\documentclass[sigconf]{acmart}
\usepackage{algorithm}
\usepackage{algpseudocode}

\usepackage{listings}
\usepackage{xcolor}
\usepackage{tabularx}   
\usepackage{booktabs}
\usepackage{tikz}
\usepackage{pgfplots}
\pgfplotsset{compat=1.18}  
\usepgfplotslibrary{statistics}
\usetikzlibrary{pgfplots.statistics}
\usepgfplotslibrary{groupplots}
\usepackage{array} 
\usepackage{float}
\usepackage{paralist}

\lstdefinelanguage{PostgreSQL}{
  keywords={
    SELECT, DISTINCT, FROM, LEFT, JOIN, USING, WHERE, AND, OR, NOT, ANY, ALL
  },
  sensitive=false,
  morecomment=[l]--,
  morecomment=[n]{/*}{*/},
  morestring=[b]',
  morestring=[b]"
}
\AtBeginDocument{%
  }

\setcopyright{acmlicensed}

\copyrightyear{2026}
\acmYear{2026}
\acmDOI{10.1145/3748522.3779850}
\acmConference[SAC '26]{The 41st ACM/SIGAPP Symposium on Applied Computing}{March 23--27, 2026}{Thessaloniki, Greece}
\acmISBN{979-X-XXXX-XXXX-X/26/03}

\begin{document}


\title{Aurora: Neuro-Symbolic AI Driven Advising Agent}


\author{Lorena Amanda Quincoso Lugones}
\email{lorena.a.quincoso@gmail.com}
\affiliation{%
  \institution{Florida International University}
  \city{Miami}
  \state{FL}
  \country{United States}
}

\author{Christopher Kverne}
\email{ckver001@fiu.edu}
\affiliation{%
  \institution{Florida International University}
  \city{Miami}
  \state{FL}
  \country{United States}
}

\author{Nityam Sharadkumar Bhimani}
\email{nick.bhimani@outlook.com}
\affiliation{%
  \institution{Northeastern University}
  \city{Boston}
  \state{MA}
  \country{United States}
}

\author{Ana Carolina Oliveira}
\email{anamalveira098@gmail.com}
\affiliation{%
  \institution{Florida International University}
  \city{Miami}
  \state{FL}
  \country{United States}
}

\author{Agoritsa Polyzou}
\email{apolyzou@fiu.edu}
\affiliation{%
  \institution{Florida International University}
  \city{Miami}
  \state{FL}
  \country{United States}
}

\author{Christine Lisetti}
\email{lisetti@cs.fiu.edu}
\affiliation{%
  \institution{Florida International University}
  \city{Miami}
  \state{FL}
  \country{United States}
}

\author{Janki Bhimani}
\email{jbhimani@fiu.edu}
\affiliation{%
  \institution{Florida International University}
  \city{Miami}
  \state{FL}
  \country{United States}
}

\renewcommand{\shortauthors}{Quincoso Lugones et al.}

\begin{abstract}

Academic advising in higher education is under severe strain, with advisor-to-student ratios commonly exceeding 300:1. These structural bottlenecks limit timely access to guidance, increase the risk of delayed graduation, and contribute to inequities in student support. We introduce \textit{Aurora}, a modular neuro-symbolic advising agent that unifies retrieval-augmented generation (RAG), symbolic reasoning, and normalized curricular databases to deliver policy-compliant, verifiable recommendations at scale. Aurora integrates three components: (i) a Boyce-Codd Normal Form (BCNF) catalog schema for consistent program rules, (ii) a Prolog engine for prerequisite and credit enforcement, and (iii) an instruction-tuned large language model for natural-language explanations of its recommendations. To assess performance, we design a structured evaluation suite spanning common and edge-case advising scenarios, including short-term scheduling, long-term roadmapping, skill-aligned pathways, and out-of-scope requests. Across this diverse set, Aurora improves semantic alignment with expert-crafted answers from 0.68 (Raw~LLM baseline) to 0.93 (+36\%), achieves perfect precision and recall in nearly half of in-scope cases, and consistently produces correct fallbacks for unanswerable prompts. On commodity hardware, Aurora delivers sub-second mean latency (0.71\,s across 20 queries), approximately $83\times$ faster than a Raw~LLM baseline (59.2\,s). By combining symbolic rigor with neural fluency, Aurora advances a paradigm for accurate, explainable, and scalable AI-driven advising.


\end{abstract}



\maketitle

\section{Introduction}
High-quality academic advising plays a vital role in supporting student success, yet it often fails to achieve that goal on many campuses. National surveys~\cite{insidehigheredStudentSurvey, NSSE2021} reveal that more than 45\% of undergraduates cannot meet with their advisor as often as needed, and many report fewer than one interaction per year. In some cases, there are more than 300-600 students per advisor ~\cite{NACADA2025}, making their workload burdensome, resulting in overscheduled calendars, rushed appointments, and mounting burnout. This advising gap and inadequate guidance result in suboptimal student choices, which, in turn, delay graduation, increase attrition, and undermine trust in institutional support structures. 



\textbf{Research challenge.} 
The core challenge is to design an academic advising system that remains accurate, transparent, and scalable despite the complexity and variability of real-world curricula. However, adaptability to institutional changes is considered a key requirement and critical component of a successful advising system~\cite{bisaso2025towards}. The key research question, therefore, is how to integrate data-driven language understanding with verifiable symbolic reasoning to produce recommendations that are both explainable and policy-compliant as curricular data evolve over time. 



\textbf{Our approach.} 
To address this advising gap, we present \textbf{Aurora}: a modular, reproducible advising framework that integrates Retrieval-Augmented Generation (RAG) with symbolic reasoning. Aurora is a neuro-symbolic AI system that combines the generative flexibility of Large Language Models (LLMs) with the formal rigor of structured rule-based inference to automate complex aspects of degree planning, particularly those requiring strict adherence to prerequisite structures and dynamic curricular policies. By handling routine, policy-driven advising tasks at scale, Aurora reduces advisor workload and enables more time for personalized, high-value student interactions.

Recent developments in neuro-symbolic AI for education highlight the growing need for hybrid systems that combine the flexibility of deep learning with the transparency and rule enforcement of symbolic reasoning. Neuro-symbolic architectures have been shown to improve trustworthiness and interpretability in educational support systems by grounding neural predictions in explicit curricular or pedagogical knowledge~\cite{hooshyar2024augmenting, tato2022infusing}. Parallel work demonstrates that integrating symbolic structures into neural models can enhance student-strategy prediction and decision reliability in learning environments~\cite{shakya2021neurosymbolic}. Most recently, Hooshyar et al.~\cite{hooshyar2025responsibleAI} emphasize the importance of responsible, auditable human--AI collaboration in advising and tutoring, noting that purely neural systems struggle to guarantee policy compliance, transparency, or verifiable reasoning. Aurora builds directly on this trajectory by unifying retrieval-augmented LLMs with symbolic rule enforcement to deliver accurate, interpretable, and verifiable degree-planning support at scale.

We contribute the following:
\begin{compactitem}
\item \textbf{A modular neuro-symbolic architecture} for advising that integrates Boyce–Codd Normal Form (BCNF) normalized data storage, SWI-Prolog \cite{wielemaker2012swi} rule enforcement, and retrieval-augmented prompting with an instruction-tuned LLM.
\item \textbf{A reproducible benchmark suite} of twenty advising queries spanning short-term scheduling, long-term degree planning, skill-aligned requests, and out-of-scope prompts, each query evaluated over five runs for robust evaluation.
\item \textbf{Empirical validation} demonstrates that Aurora improves mean cosine similarity from 0.68 (instruction-tuned LLM baseline without retrieval/symbolic) to 0.93 (+36\%), achieves perfect precision and recall in about 50 \% of the in-scope queries, and sustains interactive response times as query complexity increases.
\end{compactitem}



This study examines three research questions: \textbf{RQ1 (Short-term)}, how accurately Aurora recommends next-term courses; \textbf{RQ2 (Long-term)}, how effectively it generates prerequisite-compliant roadmaps that minimize time-to-degree; and \textbf{RQ3 (Robustness)}, how reliably it handles diverse queries, including skill-aligned and out-of-scope requests, while sustaining accuracy and interactive response times.

\section{Related Work}

(i) \textit{Rule-based and static systems.} Early automated advisors were built as flowchart systems or rule-based web applications that checked prerequisites and recommended next courses from static catalogs~\cite{murray1995decision,wehrs1992using}. A system demonstrated automated prerequisite checking, but this quickly became outdated when requirements shifted \cite{marques2001design}. Related expert systems further show the brittleness of hard-coded rules and workflows \cite{engin2014rule,daramola2014implementation}. Advising requests have been classified and mapped to predefined workflows, while subsequent work enhanced engagement through dialog flow templates and real-time feedback \cite{noaman2015new,kuhail2023engaging}. \textbf{Aurora overcomes this brittleness} by storing the catalog in BCNF with explicit keys and relations so that changes propagate cleanly, and by validating recommendations in Prolog, which retains the transparency of rules without hard-coding them into fragile UI workflows.

(ii) \textit{LLM-only systems.} LLMs used as conversational advisors excel at fluency and personalization~\cite{aguila2024,khan2022can,zylowski2024evaluating}, but without verified catalog grounding, they often hallucinate courses, violate prerequisites, or exceed credit limits. Abdelhamid et al. report these issues when deploying a GPT-4-based advisor at scale \cite{Abdelhamid2025}. \textbf{Aurora eliminates these failure modes} by feeding only vetted course facts and prerequisite chains to the generator and by requiring an explicit fallback when context is insufficient, which prevents hallucinated courses and enforces catalog compliance end to end.

(iii) \textit{Hybrid retrieval and neuro-symbolic reasoning.} Retrieval augmented generation grounds model outputs in institutional data, reducing hallucinations and stale advice, and joint retriever–generator training further reduces factual errors \cite{lewis2020retrieval,guu2020retrieval}. Neuro-symbolic methods add rule enforcement and transparency to neural fluency \cite{garcez2023neurosymbolic}. Hybrid architectures that integrate graph-aware reasoning with deep learning can improve multi-semester course sequencing. Yet existing hybrids often lack modularity, rely on ad hoc encodings, or omit strict enforcement of normalization and credit constraints. \textbf{Aurora advances this pattern} with a deliberately modular pipeline where retriever, SQL router, Prolog reasoner, and generator are separate services, with declarative rules for explainable enforcement and a normalized catalog that supports consistent credit and prerequisite checks across updates.

\textit{Positioning Aurora.} Aurora unifies BCNF-normalized catalog storage, SWI-Prolog~\cite{wielemaker2012swi} reasoning over prerequisites and credit caps, and retrieval-augmented prompting~\cite{lewis2020retrieval} to an instruction-tuned LLM. To our knowledge, no prior advising system jointly achieves strict policy compliance through symbolic reasoning, grounding in normalized, update resilient catalog schemas, and modular retrieval augmented prompting for explainable, scalable degree planning. \textbf{Empirically, this design outperforms an instruction-tuned LLM baseline without retrieval or symbolic reasoning} on semantic similarity and course level precision and recall while preserving interactive response times, demonstrating practical superiority in real advising scenarios.

A parallel line of research in neuro-symbolic educational systems has demonstrated the value of combining deep learning with structured pedagogical knowledge. Recent work shows that symbolic constraints can improve the interpretability and trustworthiness of neural models in tutoring and assessment environments~\cite{hooshyar2024augmenting}. Complementary approaches infuse expert educational knowledge directly into neural architectures using attention mechanisms to support personalized learning paths~\cite{tato2022infusing}. Other studies integrate symbolic reasoning to enhance student-strategy prediction and adaptive decision-making~\cite{shakya2021neurosymbolic}. Most recently, Hooshyar et al.~\cite{hooshyar2025responsibleAI} highlight the need for hybrid human--AI systems that maintain accountability and rule compliance in high-stakes educational contexts. While these systems demonstrate the promise of neuro-symbolic designs, none target the unique combination of catalog-grounded program validation, prerequisite-graph enforcement, and retrieval-augmented planning required for academic advising. Aurora extends this line of research by operationalizing neuro-symbolic reasoning for degree planning at scale, ensuring that every generated recommendation is simultaneously valid, auditable, and aligned with institutional policies.

\section{Aurora's Architecture}

\begin{figure}[h]
  \centering
  \includegraphics[width=\linewidth]{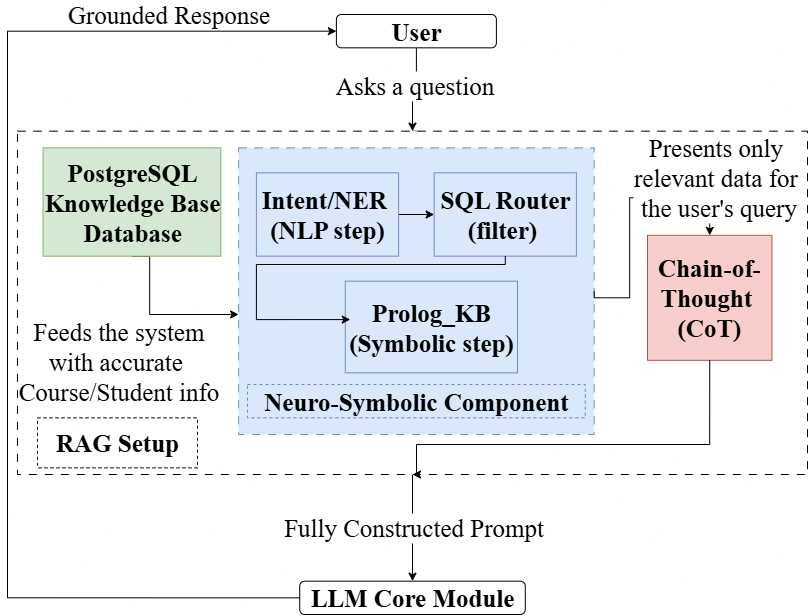}
  \caption{Aurora system overview.  Green = knowledge base; blue = neuro-symbolic reasoning; red = Chain-of-Thought prompt segment.  Solid arrows denote data flow; dashed line box show control or context boundaries.}
  \label{fig:overview}
\end{figure}

Aurora’s architecture integrates neural fluency with symbolic rigor to generate degree plans that are linguistically natural and policy-compliant. As shown in Fig.~\ref{fig:overview}, it consists of four key components: a neuro-symbolic reasoning layer, a Chain of Thought (CoT) controller, a relational database interface, and an LLM. Each module handles semantic interpretation, reasoning, validation, and generation, respectively. At the input stage, Aurora uses a dual-stage natural language understanding pipeline: Intent Recognition identifies the advising goal, and Named Entity Recognition (NER) extracts relevant entities like course codes or skills. To improve interpretability and reduce token usage, Aurora employs a structured 5W+1H prompt schema (Who, What, When, Where, Why, How) as a CoT scaffold. The knowledge base is encoded in a relational PostgreSQL schema, ensuring consistency and integrity. For the language layer, Aurora uses \textbf{DeepSeek-R1-Distill-Qwen-7B}, chosen for its fluency and lightweight design. Both the database and LLM are modular, allowing for easy substitution. The system workflow starts with the user query being processed through the intent and entity recognizers, guiding the SQL router for targeted database lookups. Retrieved data undergoes symbolic reasoning to enforce constraints, and the final output is generated with the LLM, ensuring fluency and policy compliance.



\subsection{PostgreSQL ER Knowledge Base Design}

\begin{figure}[h]
\centerline{\includegraphics[width=1\linewidth]{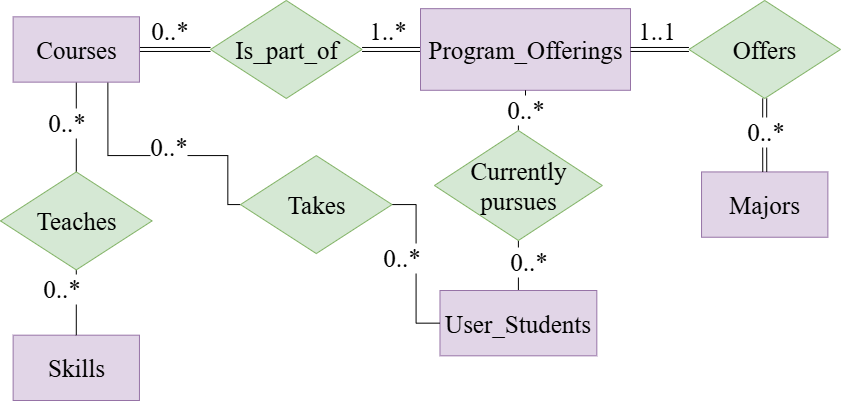}}
\caption{Conceptual ER diagram.  Purple = entities; green = relationships. Connectivity encoded in three types of links: 0..* ( e.g., a course can cover zero or many skills); 1..* (e.g., a program‐offering may include at least one course); 1..1 (e.g., every program‐offering is tied to exactly one major).}
\label{fig:er}
\end{figure}

The Conceptual ER Diagram (Fig.~\ref{fig:er}) forms the structural backbone of Aurora, linking academic and student information to enable personalized, policy-compliant advising. Aurora’s advising knowledge base is implemented as a fully normalized entity-relationship (ER) schema in PostgreSQL, connecting students, programs, courses, and skill outcomes. This schema encodes curriculum logic such as prerequisites, degree requirements, and catalog constraints within referentially sound relationships. The design ensures transparency, prevents redundancy, and guarantees consistency even as degree policies evolve. It also directly supports symbolic reasoning tasks like prerequisite traversal and program validation by exposing constraints in a relationally stable format.

\subsubsection*{PostgreSQL Consistency}\mbox{}\\
Aurora’s catalog back-end was designed directly in Boyce–Codd Normal Form (BCNF). 
We verified normalization and referential integrity through manual inspection of PostgreSQL’s internal catalog tables (\texttt{pg\_class}, \texttt{pg\_constraint}, and \texttt{pg\_attribute}). 
In every case, functional dependencies hold strictly on declared keys  for example:
\begin{itemize}
  \item \textbf{Courses:} \texttt{course\_id}~$\to$~credits, department, level, instructor, etc.
  \item \textbf{Program\_Course:} (\texttt{program\_id}, \texttt{course\_id})~$\to$~is\_core, recommended\_year.
  \item \textbf{User\_Program:} (\texttt{user\_id}, \texttt{program\_id})~$\to$~status, start\_date.
\end{itemize}
No non-key attribute was found to depend on anything other than the designated key, confirming BCNF compliance throughout the schema. 
This normalization eliminates insertion and update anomalies, ensures deterministic joins, and provides a stable foundation for retrieval and reasoning operations used in the advising workflow.

\subsection{Neuro-Symbolic Backend}

The design of Aurora’s \textbf{neuro-symbolic backend} balances flexibility and rigor. Systems that rely \emph{solely on LLMs} are  prone to violating prerequisites or credit caps, while \emph{symbolic-only engines} guarantee correctness but lack flexibility when handling the ambiguous phrasing of real student queries. Aurora  reconciles these extremes through a layered division of labor: SQL provides breadth by filtering the curriculum to a compact candidate set, Prolog supplies rigor by validating every academic constraint, and the LLM contributes fluency by phrasing compliant plans in natural language.

\subsubsection{Intent, Filtering, and Symbolic Validation}

Aurora interprets each query using lightweight intent detection and named-entity recognition to extract program identifiers, course codes, and skill targets. This structured representation guides the SQL router, which deterministically filters the catalog by removing completed courses, selecting eligible options, and applying skill, credit, and term constraints. The resulting candidate set is then passed to the Prolog engine, which enforces all hard academic rules: prerequisites, co-requisites, credit caps, and term availability. Only courses satisfying every constraint are retained. This layered design limits the LLM to a small, policy-compliant set of options, ensuring correctness while preserving flexibility in natural-language interpretation.

\subsection{Chain-of-Thought Controller and Prompting}

Aurora's Chain-of-Thought (CoT) controller converts the verified output of the symbolic backend into a compact prompt for the LLM. Instead of free-form reasoning traces, Aurora uses a structured 5W+1H template that summarizes the advising scenario through the student profile (Who), goal (What), term (When/Where), rationale (Why), and recommended actions (How). Because each field is populated directly from verified course facts and prerequisite chains, the prompt remains concise, interpretable, and grounded in catalog-consistent logic. This structure constrains the LLM to articulate verified recommendations rather than explore unconstrained reasoning paths, reducing hallucinations while preserving natural-language fluency.

Traditional chain-of-thought prompting often yields verbose or inconsistent reasoning traces that are difficult to control in educational domains.  
The 5W+1H structure offers a natural decomposition of advising intent:  
\emph{Who} represents the student profile,  
\emph{What} the target goal or requirement,  
\emph{When} and \emph{Where} the academic term and program context,  
\emph{Why} the curricular rationale (e.g., prerequisites or degree milestones), and  
\emph{How} the recommended action plan.  
Because these fields map directly to validated entities from the knowledge base, the prompt remains both interpretable and verifiable; each generated statement can be traced back to a specific symbolic fact.

\paragraph{From RAG to structured prompting.}
The retrieval-augmented context returned by the backend (\texttt{COURSE\_FACT} and \texttt{PREREQ\_CHAIN} blocks) populates the relevant 5W+1H fields automatically.  
This transformation ensures that the model receives all necessary grounding information without redundant examples or long demonstrations.  
By constraining the reasoning space in this way, Aurora reduces hallucination risk while preserving generative fluency: the LLM can focus on articulation rather than factual recall.  
The resulting prompt thus functions as a \emph{symbolically informed reasoning frame}, aligning the model’s narrative output with verified academic logic.


\subsection{LLM Core Module}
\label{sec:llm}
The final output is generated by the LLM Core Module, which uses the open, instruction-tuned \texttt{DeepSeek-R1-Distill-Qwen-7B} model. Quantized for real-time inference on consumer GPUs, the model operates under a strict system directive: respond in two sections (\texttt{<think>}, \texttt{<response>}), cite only provided evidence blocks, and emit \texttt{INSUFFICIENT\_CONTEXT} if required data is missing. Controlled decoding (beam search, low temperature) ensures factual consistency and reproducibility.

In summary, together, these modules form a robust, interpretable pipeline: the \textbf{Intent \& Named Entity Recognition (NER) service} performs precise NLP extraction; the \textbf{SQL Router} trims the search space; the \textbf{Prolog Reasoner} guarantees correctness; the \textbf{CoT Controller} scaffolds generation; and the \textbf{LLM Core} delivers natural explanations. By cleanly separating flexible interpretation from symbolic enforcement, Aurora achieves policy-compliant, student-friendly advising grounded in logic, delivered in language.



\subsection{Neuro-Symbolic Paradigm Classification}

Following the taxonomy proposed by Hooshyar et al.~\cite{hooshyar2025responsibleAI}, neuro-symbolic AI systems in education can be instantiated through several integration patterns between neural and symbolic components (e.g., Symbolic$\rightarrow$Neuro, Symbolic[Neuro], Neuro$\rightarrow$Symbolic, Neuro[Symbolic]). These paradigms differ in how knowledge is represented, where learning occurs, and whether symbolic rules are embedded inside neural networks or kept as external reasoning engines. Hooshyar et al.\ argue that modular designs, in which neural and symbolic subsystems remain separate but exchange intermediate representations, are especially promising for responsible, auditable educational AI~\cite{hooshyar2025responsibleAI,hooshyar2024augmenting,tato2022infusing,shakya2021neurosymbolic}.

Aurora follows this \textit{hybrid modular neuro-symbolic pipeline} pattern. The neural layer provides semantic interpretation and natural-language generation (NER, embeddings, LLM reasoning), while the symbolic layer enforces institutional rules through SQL filtering and Prolog-based prerequisite and credit-cap validation. The symbolic subsystem constrains the search space \emph{before} neural generation, guaranteeing policy compliance without modifying the underlying model weights. Unlike symbolic-to-neural or neural-to-symbolic integration, Aurora does not embed rules inside the LLM or attempt to learn catalog logic from data; instead, it composes verified symbolic outputs with neural fluency to achieve reliable and interpretable degree-planning recommendations. This places Aurora squarely within the modular neuro-symbolic paradigm advocated for trustworthy educational AI systems~\cite{hooshyar2025responsibleAI}.

\subsection{Limitations of LLMs and Aurora’s Mitigations}

Large language models exhibit well-known limitations, including hallucination, bias, and opaque internal reasoning. Left unconstrained, they may invent courses, misinterpret catalog policies, or over-generalize from unrelated training data. Aurora mitigates these risks through a strict neuro-symbolic architecture in which the LLM never operates on unverified information: all course facts, prerequisite chains, and credit constraints are produced by SQL and Prolog before generation. The system directive prohibits the model from using external knowledge and requires an explicit \texttt{INSUFFICIENT\_CONTEXT} token when validated evidence is missing, preventing unsupported recommendations. Structured 5W+1H prompting further limits the LLM to articulating verified reasoning rather than generating free-form chains-of-thought. Although Aurora cannot eliminate the intrinsic biases of the underlying model, it confines generation to a symbolically validated search space, reducing opportunities for biased or hallucinated outputs. These safeguards convert the LLM from a black-box decision-maker into a natural-language interface layered on top of transparent, auditable symbolic reasoning.

\section{Experimental Setup}
\label{sec:evaluation}

\subsection{Design Rationale}
Evaluating an advising system requires ensuring both rule-compliance and practical usefulness. Because real advising logs require IRB approval and introduce uncontrolled behavioral factors, this study uses a controlled simulation to isolate internal reasoning accuracy under realistic curricular constraints.

\subsection{Data and Simulation Environment}
We built a reproducible synthetic advising environment covering four representative programs (CS, Data Science, IT, Product Management). Each simulated profile included academic history, degree requirements, and credit limits. Program data were sourced from the institutional catalog, and course–skill mappings followed common taxonomies. While synthetic data lack behavioral realism, they support deterministic benchmarking and isolate reasoning accuracy from human variability.

\subsection{Benchmark and Ground Truth}
To probe Aurora’s reasoning breadth, we designed a 20-query benchmark spanning four advising contexts:

\begin{itemize}
    \item Short-term scheduling: Selecting next-semester courses given completed history and credit caps.
    \item Long-term roadmapping: Constructing prerequisite-compliant degree plans.
    \item Skill-aligned planning: Generating domain-specific tracks (e.g., AI-oriented schedules).
    \item Out-of-scope recognition: Detecting and declining queries beyond academic policy.
\end{itemize}

Table \ref{tab:query-samples} shows samples of questions from each category of questions. Each query was paired with an expert reference answer produced independently by two senior students serving in advisor roles, reconciled through consensus. This provided a consistent ground truth for evaluating both factual correctness and explanation quality.

\subsection{Metrics and Experimental Setup}
Performance was assessed along three  complementary dimensions. {\em Semantic alignment:} cosine similarity between sentence-embedding vectors of Aurora’s output and expert responses, serving as a proxy for conceptual coherence and ordering of recommendations. {\em Course-level accuracy:} precision, recall, and F$_1$ scores for recommended courses relative to ground truth. {\em Operational efficiency:} mean runtime per query. These metrics jointly capture whether Aurora (i) reasons consistently with expert logic, (ii) adheres to catalog constraints, and (iii) remains computationally practical for real-time use. Experiments were run on a commodity workstation (Intel i9-14900KF CPU, RTX 4070 Ti GPU, 32 GB RAM) to mirror small university-lab conditions. The baseline condition used the same instruction-tuned LLM (DeepSeek-R1-Distill-Qwen-7B)  without retrieval or symbolic reasoning, referred to as \textbf{Raw-LLM} henceforth. All model parameters were held constant. This isolates the contribution of Aurora’s neuro-symbolic pipeline from model or hardware variation.

\subsection{LLM Baseline Protocol (Runs \& Scoring)}
For comparability, the Raw-LLM baseline used the same instruction-tuned model and decoding settings as Aurora’s generator, but without retrieval or symbolic reasoning. Every benchmark query was executed across multiple runs on the same hardware, and performance was computed from the extracted course IDs in the model’s response (course-level precision, recall, F1), plus cosine similarity of the narrative against the expert rationale, and mean response time per query. This isolates the effect of \textit{neuro-symbolic grounding} (retrieval + SQL + Prolog) from model or hardware differences. 

\subsection{Implementation details and traceability}

\subsubsection{SQL filtering}
The SQL router trims candidate lists prior to symbolic validation. The exact filter used during experiments is shown in Listing~\ref{lst:filter_query}.

\begin{lstlisting}[language=SQL,caption={Filter by program, skill, credit, and taken list}, label={lst:filter_query}]
SELECT DISTINCT pc.course_id
FROM   program_course pc
LEFT   JOIN course_skill cs USING(course_id)
LEFT   JOIN courses      c  USING(course_id)
WHERE  pc.program_id = %(program)s
  AND (%(has_skills)s AND cs.skill_id = ANY(%(skills)s)
       OR NOT %(has_skills)s)
  AND (%(has_cap)s    AND c.credits <= %(cap)s
       OR NOT %(has_cap)s)
  AND pc.course_id <> ALL(%(taken)s);
\end{lstlisting}

\subsubsection{Prolog knowledge base and greedy planner}
Aurora's hard-constraint checker runs in SWI-Prolog. For roadmaps Aurora uses a greedy planner (see Algorithm~\ref{alg:planner}) which packs eligible courses by an "unlock weight" heuristic across semesters; the planner avoids cycles because curricula prerequisites form a DAG.

\begin{algorithm}[h]
\caption{Greedy roadmap planner (high level)}
\label{alg:planner}
\begin{algorithmic}[1]
\Require \textit{Program}, \textit{Taken}, \textit{CreditCap}, \textit{Start}
\State $Need \gets$ all remaining program courses
\State $Semester \gets$ \textit{Start}; \quad $Plan \gets [\,]$
\While{$Need\neq\emptyset$}
\State $Cap \gets$ credit limit for $Semester$
\State $Elig \gets {c\in Need \mid prereqs\_met(c,Taken)}$
\If{$Elig=\emptyset$} \Comment{nothing unlocked}
\State $Seed\gets$ first course in $Need$
\State $Pick\gets$ $Seed$ plus just enough of its prerequisite \emph{unlockers} to fit $Cap$
\Else \Comment{some courses unlocked}
\State rank $Elig$ by \emph{unlock weight} = \# of future courses each unlocks
\State greedily pack highest-weight courses up to $Cap$ into $Pick$
\If{$Pick=\emptyset$} \State $Pick\gets\{\text{FIRST}(Elig)\}$ \EndIf
\EndIf
\State add co-requisites to $Pick$
\State add any missing prerequisites or allowed alternatives
\State trim or pad $Pick$ to respect $Cap$ and the three-course minimum
\State append block$(Semester,Pick)$ to $Plan$
\State $Taken\gets Taken\cup Pick$; $Need\gets Need\setminus Pick$
\State $Semester\gets$ next academic term
\EndWhile
\State \Return $Plan$
\end{algorithmic}
\end{algorithm}

\subsubsection{Prompt assembly and 5W1H framing}
After Prolog certifies a candidate set, the router serializes evidence (STUDENT\_HISTORY, COURSE\_FACT blocks, optional PREREQ\_CHAIN), inserts a compact 5W+1H header, and hands the resulting prompt to the generator. An anonymized example prompt body used in evaluation is shown in Listing~\ref{lst:prompt_body}.

\begin{lstlisting}[ float,
  floatplacement=tb,
  basicstyle=\ttfamily\scriptsize,
  caption={Illustrative prompt body (anonymized).},
  label={lst:prompt_body}]
### STUDENT_QUERY 'I would like a machine-learning schedule next spring, max 12 credits.'
### STUDENT_HISTORY ABC1010 DEF2020 GHI3030
### COURSE_FACT id = MLA4100 name = Intro to Machine Learning credits = 3 description = 'Supervised and unsupervised basics' id = DST3300 name = Data-Science Tools credits = 3 description = 'Python, arrays, data frames.' . . . (additional vetted courses in the same format)
### PREREQ_CHAIN MLA4100 <- GHI3030, DEF2020 DST <- ABC1010
### 5W1H FRAME Who: B.S. Computer Science What: machine-learning schedule next spring, max 12 credits When: Spring 2026 Where: n/a Why: machine learning, data science How: using the vetted courses above
\end{lstlisting}

\vspace{1em}

\subsubsection{Model configuration and system message}
Aurora's narrative layer used \texttt{DeepSeek-R1-Distill-Qwen-7B} quantized with 4-bit NF4; generation employed a conservative beam-decoding profile. The fixed system directive enforces the two-block output contract and mandates \texttt{INSUFFICIENT\_CONTEXT} when necessary. An excerpt from the system message is shown in Listing~\ref{lst:systemmsg}.

\begin{lstlisting}[basicstyle=\ttfamily\scriptsize,
caption={Excerpt from the fixed system message.}, label={lst:systemmsg}]
### System
You are *Aurora*, an academic-advising assistant [. . .]
Always output exactly two sections:
  1) <think>...</think>
  2) <response>...</response> beginning with
     'As your academic advisor, I recommend...'
Use only the supplied context [. . .]
If you need data you don't see, reply exactly:
INSUFFICIENT_CONTEXT
\end{lstlisting}

\subsubsection{Interpretability and provenance}

Every generated recommendation is accompanied by provenance: the SQL used to produce candidates, the Prolog trace that validated the set, and the evidence blocks passed to the LLM. This full-stack trace enables reproducibility, audits, and fine-grained debugging crucial for institutional adoption.

\begin{table}[b]
\small\centering
\begin{tabularx}{\columnwidth}{l l X}
\toprule
\textbf{Category} & \textbf{Persona} & \textbf{Sample Query} \\
\midrule
Short-term (Next Term)   & CS Major   & What courses should I take next semester to stay on track for my CS-BS degree? \\
Long-term (Degree Plan)  & CS Major   & Plan the rest of my CS-BS degree so I can graduate on time. \\
Skill-aligned            & CS Major   & I’d like an AI-oriented schedule for next Fall. What should I take? \\
Short-term (Next Term)   & CS Minor   & Please suggest my next-term schedule to finish the CS minor (12-credit cap). \\
Out-of-scope             & CS Major   & What can I do if a class is difficult? \\
\bottomrule
\end{tabularx}
\caption{Representative evaluation queries.}
\label{tab:query-samples}
\end{table}

\section{Results: Aurora-at-work}

\begin{figure}[b]
  \centering
  \begin{tikzpicture}
    \begin{axis}[
      width=0.6\linewidth,
      height=5cm,
      ybar,
      bar width = 3pt,
      ymin = 0, ymax = 1.05,
      ylabel = {Mean Cosine Similarity (5-cycle average)},
      ytick  = {0,.2,.4,.6,.8,1},
      symbolic x coords = {Q1,Q2,Q3,Q4,Q5,Q6,Q7,Q8,Q9,Q10,Q11,Q12,Q13,Q14,Q15,Q16,Q17,Q18,Q19,Q20},
      xtick  = data,
      xticklabel style = {rotate=45, anchor=east, font=\scriptsize},
      x=0.35cm,
      enlarge x limits = 0.05,
      grid=major, grid style={dashed,gray!30},
      clip=false,
      legend style={
        at={(axis description cs:0.5,-0.15)},
        anchor=north,
        legend columns=2,
        font=\footnotesize
      },
      legend image code/.code={
        \draw[#1,fill=#1] (0cm,-0.1cm) rectangle (0.3cm,0.2cm);
        },
      title = {Mean Cosine Similarity (5-cycle average)},
    ]

      \addplot+[draw=none, fill=blue!70]
        coordinates {
          (Q1,0.9383) (Q2,0.9450) (Q3,0.9270) (Q4,0.9577) (Q5,0.8893)
          (Q6,0.9062) (Q7,0.9752) (Q8,0.9300) (Q9,0.9894) (Q10,0.9643)
          (Q11,0.9219) (Q12,0.9865) (Q13,0.9120) (Q14,0.9208) (Q15,0.9126)
          (Q16,0.7414) (Q17,0.9236) (Q18,1.0000) (Q19,1.0000) (Q20,1.0000)
        };

      \addplot+[draw=none, fill=orange!80]
        coordinates {
          (Q1,0.8344) (Q2,0.6208) (Q3,0.6947) (Q4,0.7036) (Q5,0.5234)
          (Q6,0.5873) (Q7,0.5508) (Q8,0.5667) (Q9,0.7704) (Q10,0.8234)
          (Q11,0.6767) (Q12,0.7235) (Q13,0.5074) (Q14,0.6790) (Q15,0.5949)
          (Q16,0.7203) (Q17,0.8447) (Q18,0.2709) (Q19,0.3443) (Q20,0.2547)
        };

      \legend{Aurora (RAG), Raw-LLM (DeepSeek)}
    \end{axis}
  \end{tikzpicture}
  \caption{Per-query cosine similarity averaged over five runs.}
\label{fig:aurora_vs_raw_llm}
\end{figure}
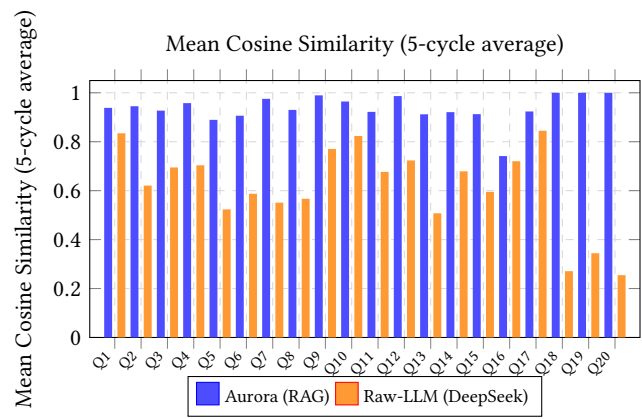

Figure~\ref{fig:aurora_vs_raw_llm} shows per-query mean cosine similarity between system outputs and ground-truth answers. Higher values indicate greater semantic alignment with handcrafted expert responses.

Across all in-scope benchmark queries, Aurora achieves a mean cosine similarity of 0.93 compared to 0.68 for the Raw-LLM baseline, indicating substantially closer alignment with expert-crafted advising recommendations. Course-level precision, recall, and F$_1$ scores average 0.81–0.83, whereas the unconstrained LLM frequently proposes ineligible or nonexistent courses, resulting in near-zero accuracy. Aurora also maintains sub-second end-to-end latency by pruning the catalog through SQL and validating prerequisites and credit constraints in Prolog before generation. This ensures that the LLM focuses solely on phrasing verified decisions rather than performing unconstrained reasoning. Out-of-scope prompts reliably trigger the fallback template, demonstrating controlled behavior and policy-compliant robustness.

The entire planning and generation pipeline completes in $<1$ s on commodity hardware (Intel i9 CPU, RTX 4070 GPU), demonstrating that precise reasoning and natural-language fluency can coexist efficiently within a unified advising architecture.

\subsection{Validation of System Consistency and Efficiency}

To confirm Aurora’s robustness and architectural soundness, we performed a two-part sensitivity analysis focusing on (i) the relational data layer and (ii) the retrieval–generation (RAG) footprint. Together, these tests verify that the system’s foundations are both theoretically consistent and practically efficient under realistic advising workloads.

\subsubsection*{RAG Footprint Analysis}\mbox{}\\
Retrieval-augmented generation (RAG) constrains Aurora’s prompt size by fetching only the minimal catalog subset relevant to each query. 
Without retrieval, the system would serialize the entire course catalog into the model prompt  approximately \(210 \times 60 \approx 12{,}600\) tokens. 
Empirically, across the benchmark suite, Aurora’s retriever produces a mean prompt size of \(492 \pm 444\) tokens per query (range: 0–1,346) and retrieves on average \(7.8 \pm 8.7\) courses (range: 0–24). 

The observed reduction ratios are:
\begin{center}
\begin{tabular}{@{}ll@{}}
Worst case & \(\dfrac{1{,}346}{12{,}600} \approx 0.11\;\text{(11\%)}\)\\[2pt]
Typical    & \(\dfrac{492}{12{,}600} \approx 0.04\;\text{(4\%)}\)
\end{tabular}
\end{center}

Simple or out-of-scope queries yield zero-length prompts, confirming that the early-exit logic is functioning as intended. 
By logging \(\langle\text{queryID},N_{\text{retrieved}},|\text{tokens}_{\text{RAG}}|\rangle\) for every inference call, Aurora maintains full auditability of retrieval behavior and prompt expansion. 
Overall, this analysis confirms that RAG consistently reduces token utilization by over an order of magnitude while preserving completeness and recall, ensuring scalability and responsiveness in real-world academic advising contexts.

\subsection{Takeaways from the Evaluation}

\textbf{Accuracy and Reliability.} 
Aurora improves semantic alignment from 0.68 to 0.93 (+36\%) and achieves perfect precision and recall on roughly half of all in-scope queries, demonstrating that symbolic grounding yields verifiably correct degree plans.

\textbf{Robustness and Policy Compliance.} 
Out-of-scope prompts consistently trigger explicit fallbacks, ensuring that uncertainty is surfaced rather than glossed over, which is critical for institutional deployment.

\textbf{Efficiency and Scalability.} 
SQL filtering and Prolog validation execute in milliseconds, keeping end-to-end latency under 1 s. Because the LLM receives only compact, validated context, generation (rather than reasoning) dominates runtime.

\textbf{Traceability and Explainability.} 
Every recommendation is fully auditable from SQL retrieval to Prolog validation and the final prompt, supporting reproducibility and future advisor oversight.

\textbf{Broader Significance.}
Taken together, these results position Aurora as a concrete demonstration that neuro-symbolic architectures can make large language models \emph{operationally trustworthy} in educational decision-making. 
By achieving near-expert accuracy, safe fallbacks, and transparent reasoning on commodity hardware, Aurora transforms LLM-based advising from a proof-of-concept into a viable institutional tool. 
More broadly, the system illustrates how symbolic grounding can turn generative models into dependable collaborators, not replacements, for human advisors, enabling scalable, explainable support for students while preserving academic oversight.

\section{Discussion}
\subsection{Interpretation of Findings}
Aurora’s evaluation demonstrates that combining symbolic reasoning with neural generation can substantially improve internal reasoning fidelity in automated academic advising. The observed gains in semantic alignment (from 0.68 to 0.93 cosine similarity) and balanced precision–recall confirm that grounding an LLM in structured curricular logic reduces factual errors and enforces prerequisite compliance. These results highlight an important methodological insight: large language models can reason more reliably when constrained by explicit, verifiable rules rather than free-text prompts alone. At the same time, these findings should be interpreted within the limits of a controlled simulation. The experiment validates the architectural integrity of Aurora: its ability to produce consistent, policy-compliant outputs, not its educational impact. Because the test data were synthetically generated and devoid of human variability, these results speak to internal validity rather than to student learning outcomes or advising satisfaction. In this sense, the work provides evidence of technical reliability, not proof of pedagogical effectiveness. The improvements over an instruction-tuned LLM baseline (Raw-LLM) also underscore a broader research point: retrieval and symbolic verification do not merely add interpretability but actively shape reasoning behavior. Aurora’s architecture thus serves as a replicable framework for other educational AI systems where correctness, auditability, and transparency are critical.


\subsection{Educational and Practical Implications}

Aurora demonstrates how neuro-symbolic methods can improve transparency and accountability in advising systems. The provenance trail (from SQL filtering to Prolog validation) supports auditability and provides advisors with clear insight into how recommendations were produced. Sub-second latency enables real-time interaction on standard hardware. Aurora is therefore best understood as a co-advising assistant: it automates rule enforcement and data retrieval, while human advisors supply contextual judgment and mentoring. Integrating Aurora into existing workflows will require interface design that surfaces explanations effectively and training that helps advisors interpret AI-supported guidance.

\subsection{Limitations and Future Directions}The present study’s primary limitation lies in its reliance on simulated student data. While this approach enabled reproducible testing of reasoning accuracy, it cannot capture the behavioral, motivational, or affective dynamics of real advising conversations. Consequently, Aurora’s high precision and recall reflect compliance with curricular logic, not demonstrable improvement in student outcomes. A second limitation is interactional realism: the current evaluation omits live user interaction and human–AI collaboration effects. Factors such as trust, interpretability, and cognitive load were not measured. Additionally, the benchmark represents a single institution’s catalog and policy framework; broader generalization across diverse universities would require adapting both schema design and rule sets. Future research should therefore pursue (i) field validation with real advisors and students, assessing not only accuracy but also usability and trust; (ii) user-interface design to enhance explainable feedback and ensure high usability based on end-user feedback, (iii) comparative studies examining how advisors interpret or override Aurora’s recommendations; (iv) expanded reasoning coverage, extending beyond prerequisite graphs to include electives, double majors, and institutional exceptions; and (v) fairness and bias analysis, ensuring equitable treatment across academic pathways.
\section{Conclusion}
This study presented Aurora, a neuro-symbolic advising system that integrates retrieval-augmented generation with rule-based reasoning to deliver verifiable, policy-compliant course recommendations. By grounding large-language-model outputs in normalized curricular data and enforcing prerequisite and credit constraints through symbolic logic, Aurora achieves high semantic accuracy and efficient, near-real-time responses. Within a controlled, simulated evaluation, Aurora outperformed an instruction-tuned LLM baseline on all major advising tasks, providing credible evidence that LLMs, when equipped with highly specialized and context-constrained reasoning layers, can produce substantially more reliable results than unconstrained generative systems. While further validation with real students and authentic advising data remains necessary, Aurora already demonstrates how verifiable neuro-symbolic architectures can move AI-driven advising from experimental prototypes toward practical, auditable tools for higher education and lay the foundation for future human-in-the-loop deployments that emphasize trust, usability, and educational impact.
\begin{acks}
This work was supported in part by the National Science Foundation under grants CSR-2402328, CAREER-2338457, CSR-2406069, CSR-2323100, and HRD-2225201. We acknowledge Dr. Gregory Murad Reis for his early insights on integrating RAG methods with LLM-driven reasoning, which helped reinforce the viability of our system design approach. The source code, dataset schema, and evaluation suite for Aurora will be publicly available at \url{https://github.com/Damrl-lab/Aurora}.
\end{acks}

\bibliographystyle{ACM-Reference-Format}
\bibliography{sections/references}

\end{document}